# Laser-thinning of MoS$_2$: on demand generation of a single-layer semiconductor

*A. Castellanos-Gomez\*, M. Barkelid, A. M. Goossens, V. E. Calado, H. S. J. van der Zant and*

*G. A. Steele\*.*

Kavli Institute of Nanoscience, Delft University of Technology, Lorentzweg 1, 2628 CJ Delft,

The Netherlands.

**a.castellanosgomez@tudelft.nl** , **g.a.steele@tudelft.nl**

Single-layer MoS$_2$ is an attractive semiconducting analogue of graphene that combines high mechanical flexibility with a large direct bandgap of 1.8 eV. On the other hand, bulk MoS$_2$ is an indirect bandgap semiconductor similar to silicon, with a gap of 1.2 eV, and therefore deterministic preparation of single MoS$_2$ layers is a crucial step towards exploiting the large direct bandgap of monolayer MoS$_2$ in electronic, optoelectronic, and photovoltaic applications. Although mechanical and chemical exfoliation methods can be used to obtain high quality MoS$_2$ single-layers, the lack of control in the thickness, shape, size, and position of the flakes limits their usefulness. Here we present a technique for controllably thinning multilayered MoS$_2$ down to a single-layer two-dimensional crystal using a laser. We generate single layers in arbitrary shapes and patterns with feature sizes down to 200 nm, and show that the resulting two-dimensional crystals have optical and electronic properties comparable to that of pristine exfoliated MoS$_2$ single layers.

Among two-dimensional crystals obtained by exfoliation of layered 3D materials,[1] graphene is by far the most studied due to its outstanding mechanical[2] and electronic properties.[3] Other 2D crystals, however, have recently gained considerable interest, since their properties are complementary to those of graphene.[4-8] For instance, the lack of a bandgap in graphene, which yields small current on/off ratios in graphene-based field effect transistors (FETs), has motivated research in other 2D semiconductor crystals such as MoS$_2$, which possess a large intrinsic bandgap.[9]

While in its bulk form, MoS$_2$ is an indirect gap semiconductor with a 1.2 eV bandgap; monolayer MoS$_2$ on the other hand has a direct gap of 1.8 eV.[10] This indirect-to-direct transition, arising from quantum confinement effects as the thickness decreases,[11] results in an enhancement of the photoluminescence of monolayer MoS$_2$ with respect to the multilayered counterpart.[12-15] Additionally, transistors based on single-layer MoS$_2$ present large in-plane mobilities (200-500 cm$^2$V$^{-1}$s$^{-1}$) and high current on/off ratios (exceeding 10$^8$)[16] which make this material of great interest for electronic devices and sensors.[17-19] The excellent mechanical properties of $MoS_2$ also suggest its prospective use in flexible semiconducting applications.[20, 21]

Until now, production of monolayer $MoS_2$ has been mainly performed by mechanical and chemical exfoliation of single layers from bulk crystals.[1, 4, 12, 22] Although these exfoliation methods have proven to be effective to obtain high-quality $MoS_2$ single layers suitable for fundamental research, future applications require the development of new procedures to allow "on-demand" fabrication/modification of this nanomaterial.

Here, we present a top-down approach to fabricate $MoS_2$ single layers based on laser-thinning of multilayered $MoS_2$ flakes. The fabricated $MoS_2$ monolayers have been characterized by optical microscopy, atomic force microscopy (AFM) and Raman spectroscopy. In order to gain a deeper insight into their semiconducting properties, their photoluminescence (PL) spectra and their performance as a channel in FETs were explored. We find that laser-thinning of multilayered $MoS_2$ provides a reliable method to fabricate $MoS_2$ single layers with user-defined shape and size and with optical and electronic properties that are comparable to those of pristine $MoS_2$ single layers. The technique presented here offers the possibility to scale up the fabrication of single layer $MoS_2$.

In a typical experiment, a multilayered $MoS_2$ flake was deposited onto a $Si/SiO_2$ (285 nm) substrate by mechanical exfoliation[1] and characterized by a combination of optical microscopy, AFM and Raman spectroscopy (see Materials and Methods section). Figure 1(a) shows an optical micrograph of a multilayered $MoS_2$ flake on a $Si/SiO_2$ substrate. The regions with different color correspond to zones of the flake with different number of layers[23, 24], with the faint purple region in the center of Figure 1(a) corresponding to a single layer $MoS_2$.

A scanning laser from a confocal Raman microscope (Renishaw in via RM 2000, $\lambda = 514$ nm) was then used to thin the multilayered $MoS_2$ down to a monolayer by moving the laser over the flake with high power (see Materials and Method section for more details), in a similar way to the procedure employed to locally oxidize graphene or reduce graphene oxide[25-27]. We decided to employ a Raman system to perform the laser-thinning because it allows us to characterize *in situ* the Raman spectra and the optical contrast before and after the thinning procedure. This thinning procedure, however, can be implemented in other setups such as laser pattern generators which are optimized to perform fast laser-based lithography.

Figure 1(b) shows an optical micrograph of the same flake as in Figure 1(a) after the laser-thinning process in the region marked by the dotted rectangle in Figure 1(a). The optical contrast of the thinned region is uniform and consistent with that of a single $MoS_2$ monolayer[23, 24]. The topography of the region thinned with the laser has been studied with AFM (Figure 1(c)). As can already be inferred from the contrast in the optical image, the initially multilayered $MoS_2$ flake has not been completely removed by the laser. Instead, a single layer of material with a thickness of $0.9 \pm 0.3$ nm remains on the surface, a value consistent with the expected thickness from a $MoS_2$ monolayer.

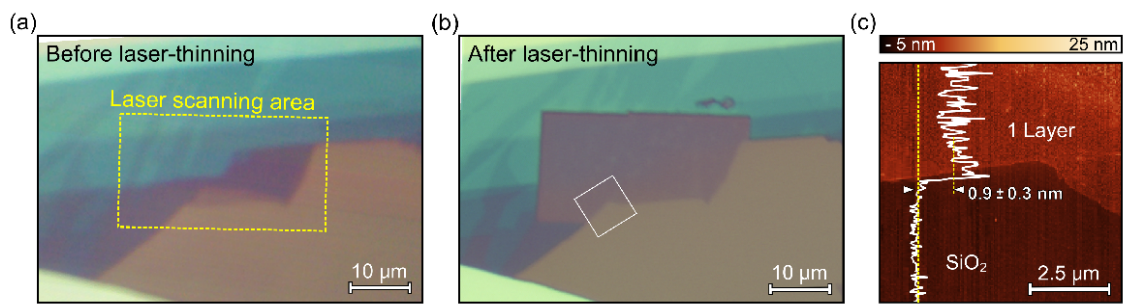

**Figure 1.** (a) Optical microscopy image of a multilayered $MoS_2$ flake deposited onto a 285 nm $SiO_2$/Si substrate. (b) Same as in (a) after scanning a laser in the area marked by a dashed rectangle in (a). The laser-thinning parameters were $\lambda$ = 514 nm, incident power on the sample 10 mW, scan step 400 nm and exposure time of 0.1 sec between steps. (c) Topographic AFM images of the region marked by the square in (b). A vertical topographic line profile is included in (c) to indicate the thickness of the laser-thinned layer.

The thinning procedure relies on the sublimation of the upper layers due to the heating induced by light absorption. Apparently the heat cannot be easily dissipated through the substrate because of the poor coupling between the $MoS_2$ layers mediated by van der Waals forces. The bottom layer, however, remains on the substrate until much higher laser powers because it is in intimate contact with the $SiO_2$/Si substrate which acts as a heat sink. Using our experimental setup, the laser-thinning method can produce 8 $\mu m^2$/min. of $MoS_2$ monolayer and thus it can be used to thin down large-area multilayer $MoS_2$ crystals as those grown by recently developed vapor-phase methods[28, 29]. The roughness of the laser-fabricated layers, however, is about three times larger than that of a pristine $MoS_2$ monolayer, probably due to the presence of unremoved $MoS_2$ traces on the surface, something we expect can be optimized by adjusting the laser scanning parameters. In the following, we characterize these laser-fabricated $MoS_2$ monolayers, demonstrating that their optical and electronic properties are comparable to that of pristine single layers.

Similar to the case of graphene, Raman spectroscopy can be employed to characterize the thickness of $MoS_2$ nanolayers[13, 30]. Using a low-power laser (<1 mW to avoid laser-heating effects[31]), the Raman spectra of the initial multilayered $MoS_2$ flakes have been measured (Figure 2(a)). As reported by Lee *et al*.[30], the frequency difference between the two most prominent Raman peaks depends monotonically on the number of $MoS_2$ layers (blue circles in Figure 2(b)). Figure 2(b) also compares the Raman spectra measured for a $MoS_2$ flake before (4 layers thick) and after laser-thinning. The intensity of both the $E^1_{2g}$ and $A_{1g}$ peaks and their frequency difference are drastically reduced after the thinning process, which is in agreement with a reduction of the thickness. The frequency difference value, however, is slightly larger than the one measured for a pristine $MoS_2$ monolayer (red square in Figure 2(b)). We attribute this discrepancy to the presence of traces of unremoved $MoS_2$ on the surface of the laser-fabricated monolayer, which would slightly modify the frequency of the Raman modes. Indeed, Eda *et al*.[12] observed a similar effect in $MoS_2$ monolayers fabricated by chemical methods and they attributed the discrepancy in the Raman mode frequencies to local thickness variations.

Additionally, the absence of a peak around 820 cm$^{-1}$ (fingerprint associated with the oxidation of MoS$_2$ and the formation of MoO$_3$) in the Raman spectra serves as a first indication to point out that the MoS$_2$ surface is not affected by oxidation during the laser-thinning process (see Supporting Information)[32]. Additionally, photoluminescence and electronic transport measurements (described in the following) support the interpretation that laser fabricated MoS$_2$ monolayers presents a semiconducting and not an insulating behavior (as it would be expected for MoO$_3$).

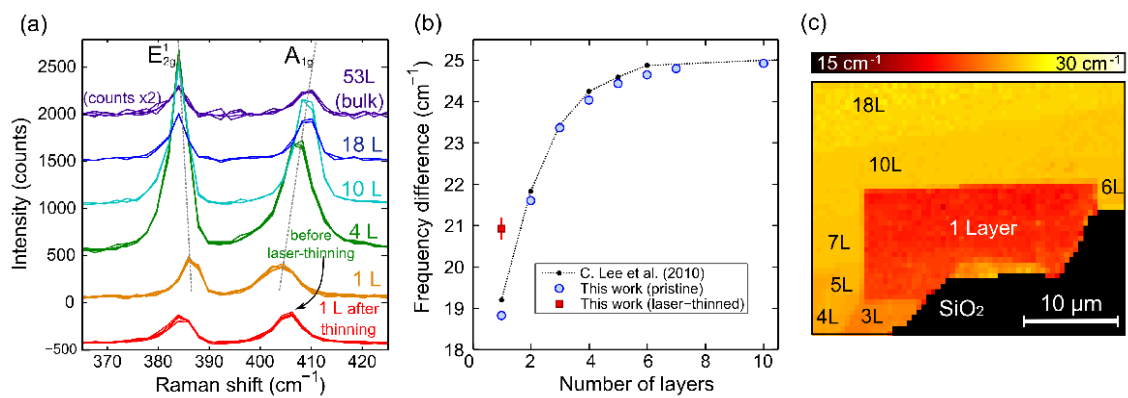

**Figure 2.** (a) Raman spectra of pristine MoS$_2$ flakes with thicknesses ranging from one single layer (1L) to 53 layers (53L). The spectrum measured for a laser-fabricated MoS$_2$ single layer is shown below the spectra for a pristine monolayer to facilitate the comparison. (b) Frequency difference between the E$^1_{2g}$ and A$_{1g}$ Raman modes. The red square represents the frequency difference measured for six laser-thinned MoS$_2$ monolayers. (c) Spatial map of the frequency difference between the Raman modes E$^1_{2g}$ and A$_{1g}$ measured in the laser-thinned flake shown in Figure 1b.

In addition to the Raman shift, single layer MoS$_2$ also exhibits a unique signature in its optical spectrum due to the transition from an indirect to a direct-bandgap semiconductor[10, 12-15]. In Figure 3, we present the optical characteristics of our laser thinned MoS$_2$ layers. The PL signal of the initial 4-6 layers thick MoS$_2$ flake is rather weak in comparison to the PL spectrum of the pristine monolayer, in agreement with previous reports[10, 15]. The PL spectrum of the laser-fabricated monolayer, however, resembles in intensity and frequency the one of the pristine MoS$_2$ monolayer, demonstrating that the laser-fabricated single layer MoS$_2$ also presents a large intrinsic direct bandgap (~1.8 eV), just as has been observed in mechanically exfoliated MoS$_2$ monolayers[10, 15]. Figure 3(b) shows a false color image of the integrated PL signal measured in a region containing thinned and pristine zones. The laser-fabricated single layer MoS$_2$ can be clearly identified by the much more intense PL signal, shown as a brighter color in the image.

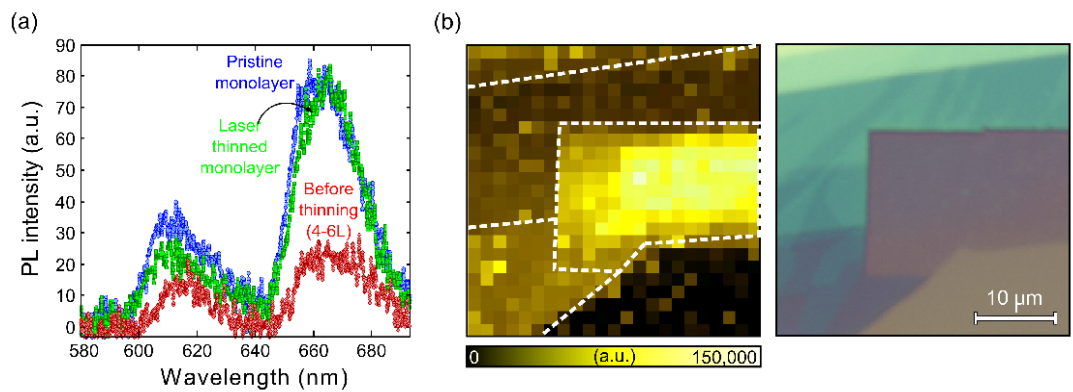

**Figure 3.** (a) Photoluminescence measured for a pristine monolayer, and for a 4-6 layers thick flake before and after laser-thinning. (b) A spatial map of the integrated photoluminescence signal between 580 nm and 700 nm, measured in the region indicated by the optical micrograph at the right. Dashed lines in PL image are a guide to the eye (extracted from the optical image shown on the right hand side) to identify the regions with different thicknesses in the $MoS_2$ flake.

To further compare the electrical properties of laser-fabricated and pristine $MoS_2$ monolayers, 11 FET devices (6 with laser-fabricated and 5 with pristine monolayers) were fabricated and characterized in two-terminal configuration. Figure 4(a) and 4(d) shows the optical micrographs of a pristine and a laser-fabricated single layer $MoS_2$ FET device respectively. The measured $I_{ds}$-$V_g$ characteristics for both pristine and laser-fabricated samples (Figure 4(b) and 4(e)) are typical of an n-doped $MoS_2$ FET and in agreement with previous reports[17, 24, 33, 34]. The current on/off ratio of both pristine and laser-fabricated devices is typically above $10^3$ (for $V_g$ ranging from -10 to 40 V), significantly higher than for graphene transistors. Figure 4(c) and (f) show the $I_{ds}$-$V_{ds}$ characteristics for the pristine and laser-fabricated FETs at different gate voltages. While pristine monolayer-based FETs exhibit a saturating behavior, the current of laser-fabricated MoS2 FETs does not saturate within the bias window studied here. It is important to note that similar non-saturating $I_{ds}$-$V_{ds}$ characteristics have also been reported for pristine monolayer $MoS_2$-based FETs[17, 24, 33, 34]. The origin of this non-saturating behavior is unknown.

The carrier mobility of the devices has been calculated from the following formula:

$$\mu = \frac{L}{W} \times \frac{d}{\varepsilon_r \varepsilon_0 \times V_{ds}} \times \frac{\partial I_{ds}}{\partial V_g} \qquad (1)$$

where $L$ is the channel length, $W$ is the channel width, $\varepsilon_0$ is $8.854 \cdot 10^{-12}$ F·m$^{-1}$, $\varepsilon_r$ is 3.9 for $SiO_2$ and $d$ is the $SiO_2$ thickness (285 nm). We found that the mobility of laser-fabricated monolayers (0.04 – 0.49 cm²V$^{-1}$s$^{-1}$) is comparable to that of pristine-monolayer based devices (0.05 – 0.85 cm²V$^{-1}$s$^{-1}$), indicating the high quality of the laser-fabricated $MoS_2$ monolayers (see Supplementary Information for the characterization of all the fabricated devices). Although higher current on/off ratios and mobilities have been achieved in $MoS_2$ FETs with $HfO_2$-based top gates, due to the dielectric screening achieved with these high-κ gate dielectrics[16, 18], the values obtained for our devices (including those based on laser-fabricated monolayers) are similar to previous studies on exfoliated $MoS_2$ FETs with a $SiO_2$-based backgate[17, 24, 33, 34].

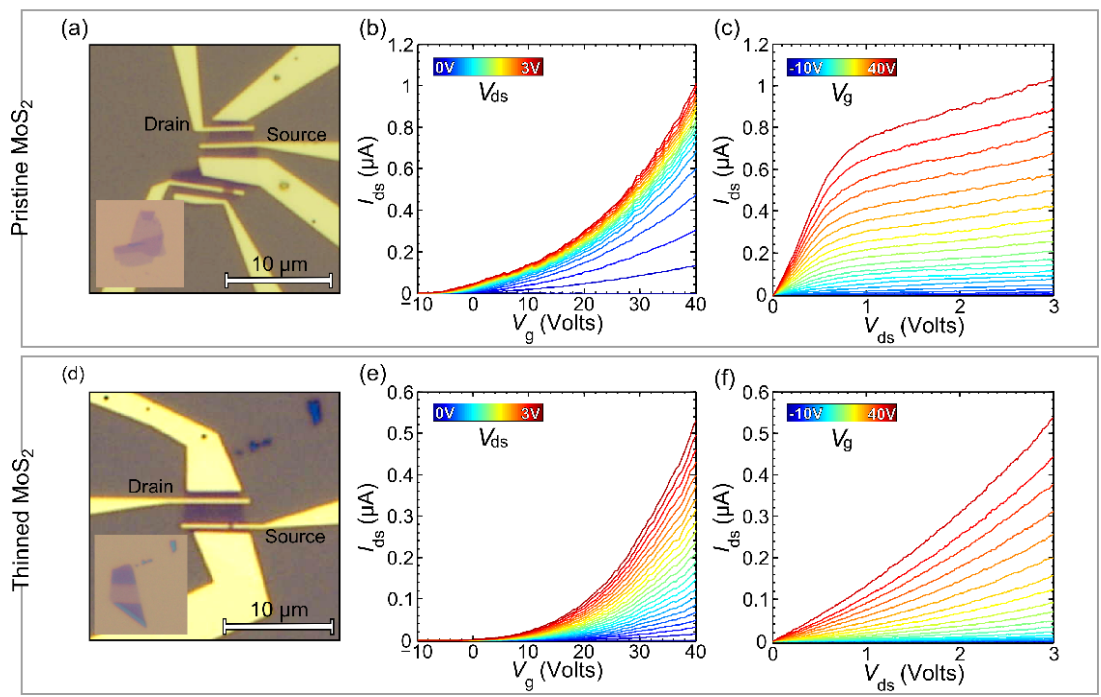

**Figure 4.** Optical micrographs of field effect transistors fabricated from a pristine $MoS_2$ monolayer (a) and a laser-fabricated monolayer (d). The insets in (a) and (d) show the $MoS_2$ flakes before processing. (b,e) Drain-source current as a function of gate voltage, measured for the pristine (b) and the laser-fabricated (e) $MoS_2$ FETs at different drain-source voltages. (c,f) Drain-source current as a function of the drain-source voltage, measure for different gate voltages, for the pristine (c) and the laser-fabricated (f) $MoS_2$ FETs.

Finally, we have further explored the possibility of employing this method to engineer the thickness of a multilayered flake following a desired pattern or even to completely cut through the flake. As the $MoS_2$ band structure depends on the number of layers, the control over the thickness allows to fabricate devices with spatially dependent electronic properties. We carried out a dose test experiment to find the optimal conditions to thin-down or to cut $MoS_2$ flakes. The scanning step size and the exposure time have been kept fixed during the dose test (400 nm and 0.1 sec, respectively) while the laser power has been used to adjust the dose. In principle, a similar dose test can be performed by keeping fixed the laser power at an intermediate power (10 mW) and changing the exposure time to adjust the dose. However, this is limited by the minimum exposure time of our experimental setup (0.1 sec). We found that scanning step size has to be of the same order of the laser spot size in order to produce homogeneous laser thinned layers.

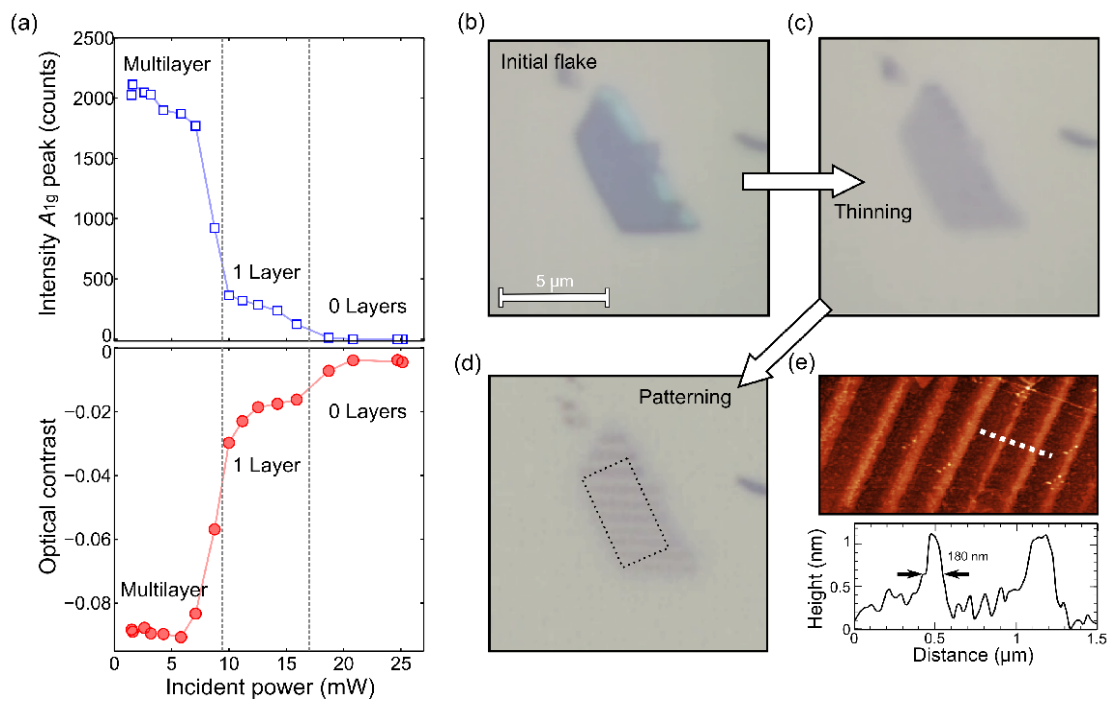

**Figure 5.** (a) Evolution of the optical contrast (bottom, red circles) and the intensity of the $A_{1g}$ Raman mode (top, blue squares) as a function of the incident laser power. These signals are used to monitor the laser-thinning/cutting of $MoS_2$ flakes. (b,c,d) Optical images of a multi-layered $MoS_2$ flake at different steps during the laser-controlled thinning and patterning process: (b) initial flake, (c) after laser-thinning, (d) after patterning it into ribbons with different widths. (e) AFM topography of the single-layer $MoS_2$ ribbons patterned by laser.

The intensity of the $A_{1g}$ Raman mode and the optical contrast[23], measured in a multilayered $MoS_2$ flake after scanning a laser with increasing power, have been employed to monitor the laser-thinning procedure. Figure 5(a) shows that while the $MoS_2$ flake remains unaffected for incident powers lower than 10 mW, a power between 10-17 mW (corresponding to a power density of 80-140 mW/µm²) thins down the $MoS_2$ flake to a single layer as can be seen from the optical contrast and the intensity of the $A_{1g}$ Raman mode. We observed a reduction of the intensity of the $A_{1g}$ Raman mode with the incident power (between 10 mW and 17 mW) which could be due to a degradation of the quality of the $MoS_2$ or to a reduction of the amount of unremoved $MoS_2$ traces on the surface while the power is increased. For even higher laser power, the $MoS_2$ layer is cut by the laser as can be inferred by the absence of the characteristic $E^1_{2g}$ and $A_{1g}$ peaks in the Raman spectra and the reduced optical contrast. According to Figure 5(a), the power necessary to thin-down or to cut a multilayered $MoS_2$ flake is well-defined and thus it is possible to use the laser thinning to engineer the thickness and the shape of a $MoS_2$ flake. Figure 5(b-d) shows the different steps of the fabrication of single-layer $MoS_2$ ribbons of different widths by laser-thinning/cutting. The topography of the fabricated ribbons is shown in Figure 5(e); we have found that ribbons as narrow as 180 nm can be fabricated with this technique, demonstrating the power of this technique to fabricate user-defined micro/nano structures in single-layers of $MoS_2$.

In summary, we have developed a procedure to laser-thin multilayered $MoS_2$ deterministically down to a single layer in arbitrary shapes and patterns with the smallest feature size ~ 200 nm. From the analysis of their photoluminescent and electronic transport characteristics, we conclude that the semiconducting properties of the laser-fabricated monolayers resemble that of pristine $MoS_2$ single layers. Moreover, adjusting the laser power this technique can be employed to reliably thin or cut multilayered $MoS_2$ flakes, paving the way to "on demand" fabrication of $MoS_2$ single layers in geometries useful for future electronic, photovoltaic and optoelectronic devices.

ACKNOWLEDGMENT

This work was supported by the European Union (FP7) through the program RODIN and the Dutch organization for Fundamental Research on Matter (FOM).

# Supplementary information:

# Laser-thinning of MoS$_2$: on demand generation of a single-layer semiconductor

*A. Castellanos-Gomez, M. Barkelid, A. M. Goossens, V. E. Calado, H. S. J. van der Zant and G. A. Steele*

Kavli Institute of Nanoscience, Delft University of Technology, Lorentzweg 1, 2628 CJ Delft, The Netherlands.

a.castellanosgomez@tudelft.nl

**Contents:**

1) **Materials and methods**
2) **Sample fabrication**
3) **Scanning electron microscopy characterization**
4) **Raman spectroscopy: probing the absence of oxide**
5) **Electrical characteristics of all the fabricated MoS$_2$-based devices**

**Materials and methods**

We prepared $MoS_2$ nanosheets on $Si/SiO_2$ substrates by mechanical exfoliation of natural $MoS_2$ (SPI Supplies, 429ML-AB) with blue Nitto tape (Nitto Denko Co., SPV 224P). To ensure optical visibility of ultrathin $MoS_2$ layers, $Si/SiO_2$ wafers with a 285 nm thermally grown $SiO_2$ layer are used[1]. We located single- and few-layer $MoS_2$ sheets under an optical microscope (Olympus BX 51 supplemented with an Olympus DP25 digital camera) and estimated the number of layers by their optical contrast[1].

Atomic force microscopy and Raman spectroscopy were also performed for a more accurate determination of the number of layers. An atomic force microscope (Digital Instruments MultiMode III AFM with standard cantilevers with spring constant of 40 N/m and tip curvature <10 nm) operated in the amplitude modulation mode has been used to study the topography and to determine the number of layers of $MoS_2$ flakes previously selected by their optical contrast. A micro-Raman spectrometer (Renishaw in via RM 2000) was used in a backscattering configuration excited with a visible laser light ($\lambda$ = 514 nm) to double check the number of layers of the studied $MoS_2$ flakes[2, 3]. The spectra were collected through a 100× objective and recorded with 1800 lines/mm grating providing the spectral resolution of ~ 1 $cm^{-1}$. To avoid laser-induced modification of the samples, all spectra were recorded at low power levels P < 1 mW[4].

Laser-thinned monolayers have been fabricated by scanning the laser of a Renishaw in via RM 2000 micro-Raman spectrometer ($\lambda$ = 514 nm) over multilayered $MoS_2$ flakes. The exposition dose has been controlled by adjusting the incident laser power, the step size and the exposure time. We found that 10 mW of incident power, a 400 nm step size and 0.1 sec exposure time yield reproducible laser-thinning of $MoS_2$ flakes with initial thicknesses of 20 layers or less. We have additionally found that snake-like scans yield more uniform thinned flakes than raster-like scans.

Photoluminescence measurements have been carried out with a homemade setup that consists of a confocal microscope with a NA = 0.8 objective illuminated by a $\lambda$ = 532 nm laser beam[5]. Typical light intensities of 9.5·$10^{-5}$ $\mu W/nm^2$ (75 µW incident power and 500 nm diameter spot) are used in this work. For the scanning photoluminescence measurement (Figure 3b) the diffraction limited spot is scanned using a combination of two galvomirrors and a telecentric lens system while a photoluminescence spectrum is acquired in each pixel.

For the nanofabrication of the $MoS_2$ based FETs, electron-beam lithography (Vistec, EBPG5000Plus HR 100) was used to write in a 495K/950K double spun PMMA layer. A 5 nm / 50 nm titanium/gold layer was evaporated. Subsequently, the PMMA/TiAu layer was gently lifted off in chloroform. Previous to their electrical characterization, samples were annealed at 250°C for two hours in a $Ar/H_2$ flow (500 sccm / 100 sccm), which increased their mobility by one order of magnitude[6]. Electronic transport of the $MoS_2$ FETs has been characterized at room temperature under high vacuum (<$10^{-5}$ mBar) in a probe station.

In order to measure the optical contrast of $MoS_2$ flakes we first acquired optical micrographs with a color camera attached to the trinocular of the optical microscope of a Renishaw in via RM 2000 system. The red channel of the images have been extracted and the resulting grayscale image has been employed to calculate the optical contrast $C = (I_{flake}-I_{subs})/(I_{flake}-I_{subs})$ where $I_{flake}$ and $I_{subs}$ are the intensity measured in the $MoS_2$ flake and the $SiO_2$ substrate respectively[1, 7].

**Sample fabrication**

Initial atomically thin $MoS_2$ flakes (1-20 layers thick) have been fabricated by mechanical exfoliation on $SiO_2$/Si substrates with 285 nm $SiO_2$ to facilitate the optical identification of ultrathin flakes [1]. The topography of the selected $MoS_2$ flakes has been studied with atomic force microscopy (Digital Instruments MultiMode III AFM) operated in the amplitude modulation mode.

Laser-thinned monolayers have been fabricated by scanning a green laser (λ = 532 nm and 10 mW of incident power) in steps of 400 nm with an exposure time of 0.1 sec between steps over multilayered $MoS_2$ flakes. Original flakes can be as thick as ~12 nm (20 layers). Thicker flakes cannot be easily thinned, probably because they sink the heat in a more efficient way than ultrathin flakes. Figure S1 shows the sharp profile of the step between the pristine and the laser-thinned area in a multilayered $MoS_2$ flake. About 9 layers have been removed by the laser.

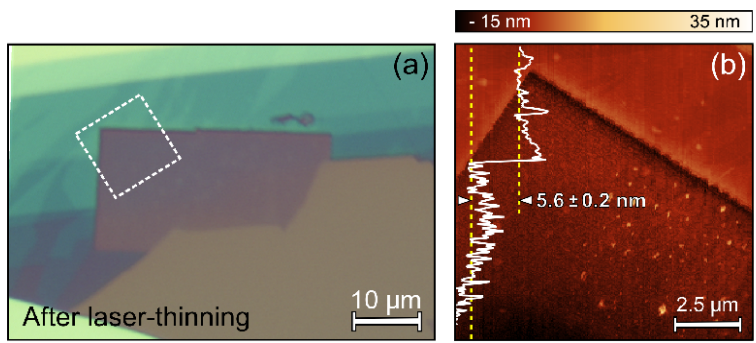

**Figure S1:** (a) Optical microscopy image of a multilayered $MoS_2$ flake deposited onto a 285 nm $SiO_2$/Si substrate after the laser thinning. (b) AFM topography of the region marked by the dashed square in (a). The boundary between the thinned and the pristine regions is show. A topographic line profile shows the sharpness of the step between the thinned and pristine zones.

Figure S2 shows optical micrographs of several multilayered $MoS_2$ flakes before and after the laser thinning procedure. The initial thickness of the different flakes is very different as can be deducted from the diversity of optical contrast.

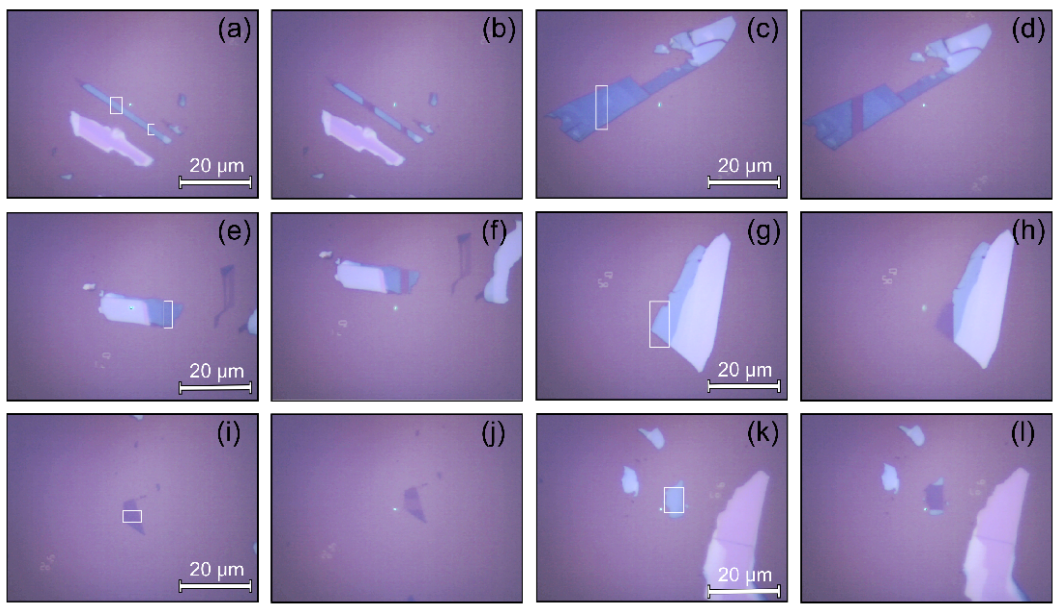

**Figure S2:** (a)/(b), (c)/(d), (e)/(f), (g)/(h), (i)/(j) and (k)/(l) are optical micrographs before/after thinning the regions marked by a white dotted rectangle.

### Scanning electron microscopy characterization

Laser-fabricated $MoS_2$ monolayers (Figure S3(a)) have been additionally characterized by scanning electron microscopy with an Hitachi ultra-high resolution FE-SEM (S-4800) (see Figure S3(b)). The laser-thinned region of the $MoS_2$ flake present a strong difference in contrast with respect to the $SiO_2$, indicating that the layer is conductive, as verified by the photoluminescence and electronic transport measurements shown in the manuscript.

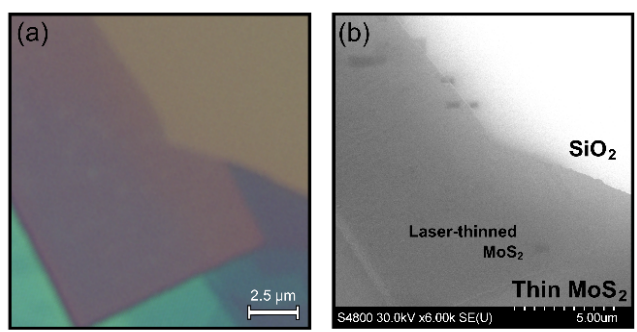

**Figure S3:** (a) Optical micrograph of the thinned $MoS_2$ flake shown in Figure 1(b) of the manuscript. (b) SEM image of the same region shown in (a).

**Raman spectroscopy: probing the absence of oxide**

In a recent work by Windom *et al.* [8] they studied the oxidation of $MoS_2$ and its fingerprints in the Raman spectra. In that work they demonstrated that $MoS_2$ is oxidized in presence of pure oxygen and high temperature and they showed that this process can be monitored by means of the presence of a peak at 820 cm$^{-1}$ related with the presence of $MoO_3$ on the surface.

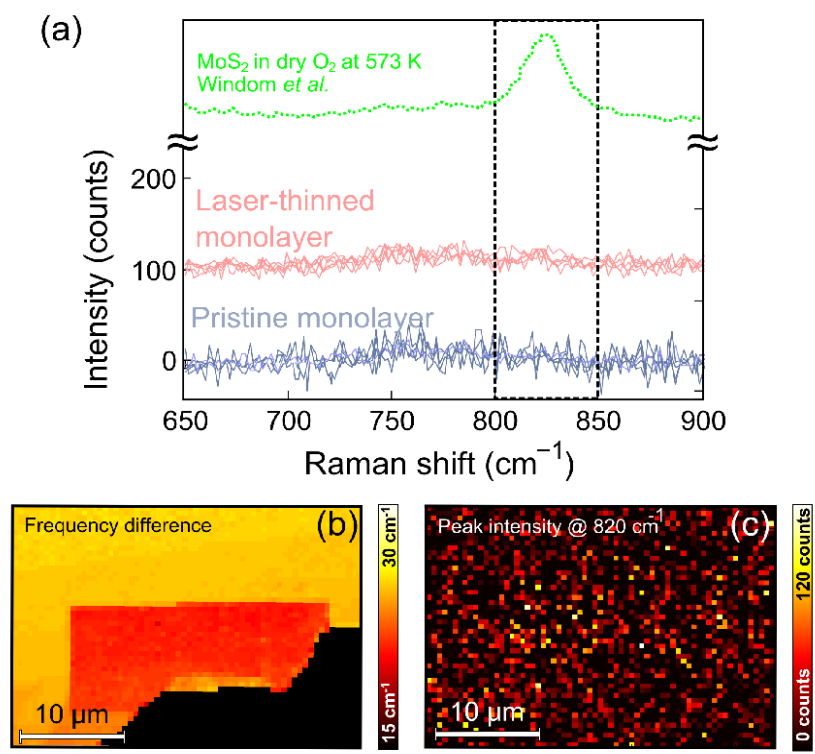

**Figure S4:** (a) Raman spectra of pristine and a laser-thinned $MoS_2$ monolayer, between 650 cm$^{-1}$ and 900 cm$^{-1}$, to probe whether the surface is oxidized or not. For comparison, it has been also added the Raman Spectrum reported for oxidized $MoS_2$ flakes in a dry $O_2$ environment at 573 K (from ref.[8] by Windom *et al*.). (b) Spatial map of the frequency difference between the $E^1_{2g}$ and $A_{1g}$ modes (shown in Figure 2(d) of the manuscript). (c) Shows a spatial map of the integrated Raman spectra between 800 cm$^{-1}$ and 850 cm$^{-1}$ to demonstrate the absence of the $MoO_3$ peak (at 820 cm$^{-1}$) in the thinned region.

Figure S4 shows a comparison between the Raman spectra measured for a pristine and a laser-thinned $MoS_2$ monolayer in the region of the spectrum in where the $MoO_3$ peak may appear. In both spectra there are no evidence of the $MoO_3$ peak which indicate that the $MoS_2$ flakes are not oxidized during the laser thinning. This is clear if we compared those spectra with the one reported in ref.[8] for $MoS_2$ oxidized in dry oxygen at 573 K that present a prominent peak at 820 cm$^{-1}$, clearly resolvable from the noise floor. Therefore, it is more probable that the process behind the laser thinning is sublimation and not burning of $MoS_2$ layers which would yield some $MoO_3$ on the surface. Figure S4(c) shows a spatial map with the integration of the Raman spectra between 800 cm$^{-1}$ and 850 cm$^{-1}$ to demonstrate the absence of the $MoO_3$ peak (at 820 cm$^{-1}$) in the thinned region.

**Electrical characteristics of all the fabricated MoS$_2$-based devices**

Characteristics of the fabricated FET devices are summarized in the following table.

| Device | Pristine / Laser-thinned | L/W | Mobility (cm$^2$V$^{-1}$s$^{-1}$) | Current on/off ratio |
|---|---|---|---|---|
| z2-fA-e12 | Pristine | 0.32 | 0.67 | 1400 |
| z5-fA-e34 | Pristine | 0.46 | 0.37 | 12404 |
| z3-fC-e12 | Pristine | 0.34 | 0.64 | 1987 |
| z3-fA-e12 | Pristine | 0.31 | 0.05 | 2601 |
| z3-fA-e45 | Pristine | 0.33 | 0.85 | 1987 |
| z3-fB-e14 | Laser-thinned | 0.27 | 0.49 | 443 |
| z3-fD-e34 | Laser-thinned | 0.25 | 0.17 | 2976 |
| z3-fD-e23 | Laser-thinned | 0.48 | 0.18 | 440 |
| z4-fD-e23 | Laser-thinned | 0.12 | 0.13 | 5347 |
| z4-fD-e12 | Laser-thinned | 0.17 | 0.04 | 2097 |
| z3-fB-e23 | Laser-thinned | 0.40 | 0.13 | 100 |

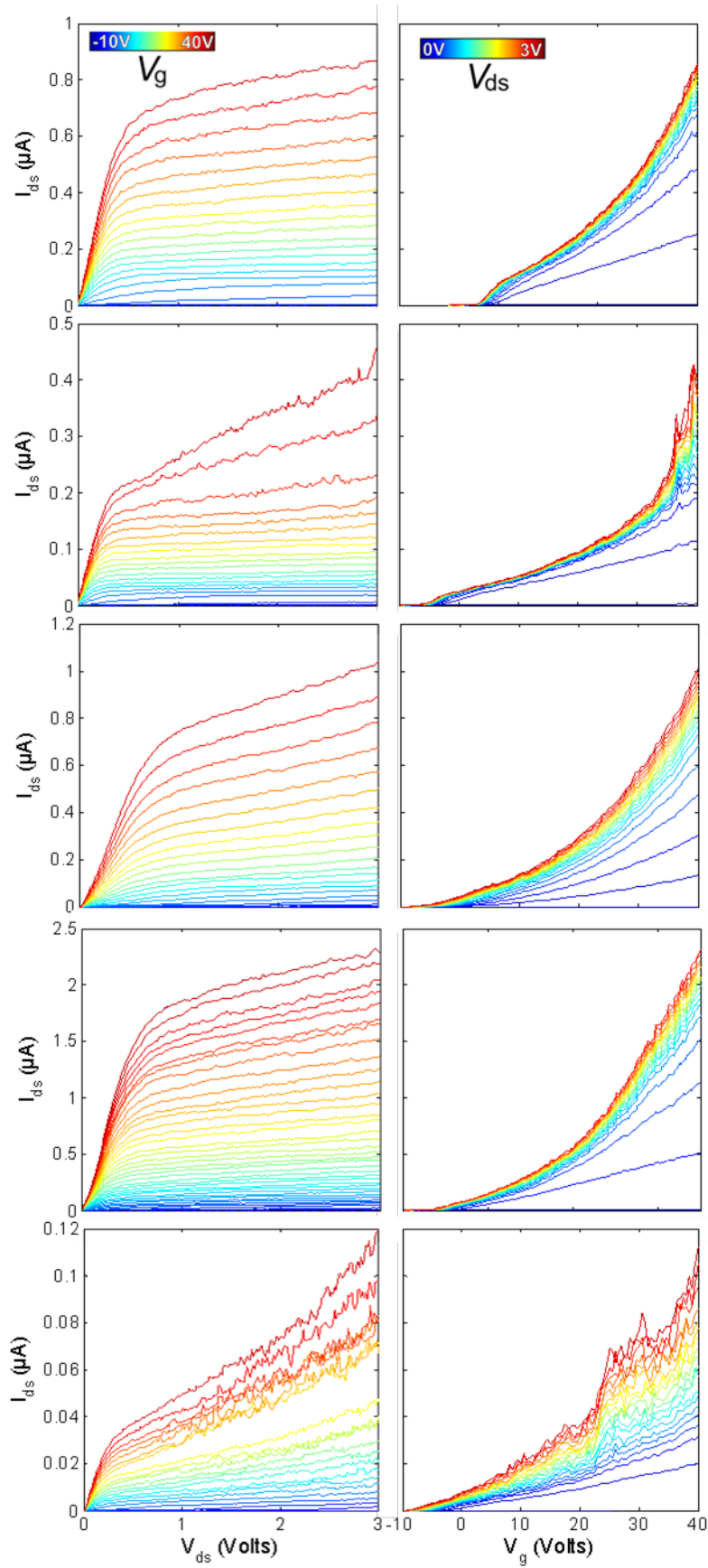

**Figure S6.** $I_{ds}$-$V_{ds}$ and $I_{ds}$-$V_g$ characteristics measured for five different FET devices based on pristine MoS$_2$ monolayers. The $I_{ds}$-$V_{ds}$ characteristics have been measured for different gate voltages ranging from -10 V (blue traces) to +40 V (red traces). The $I_{ds}$-$V_g$ characteristics have been measured for different drain source voltages ranging from -0 V (blue traces) to +3 V (red traces).

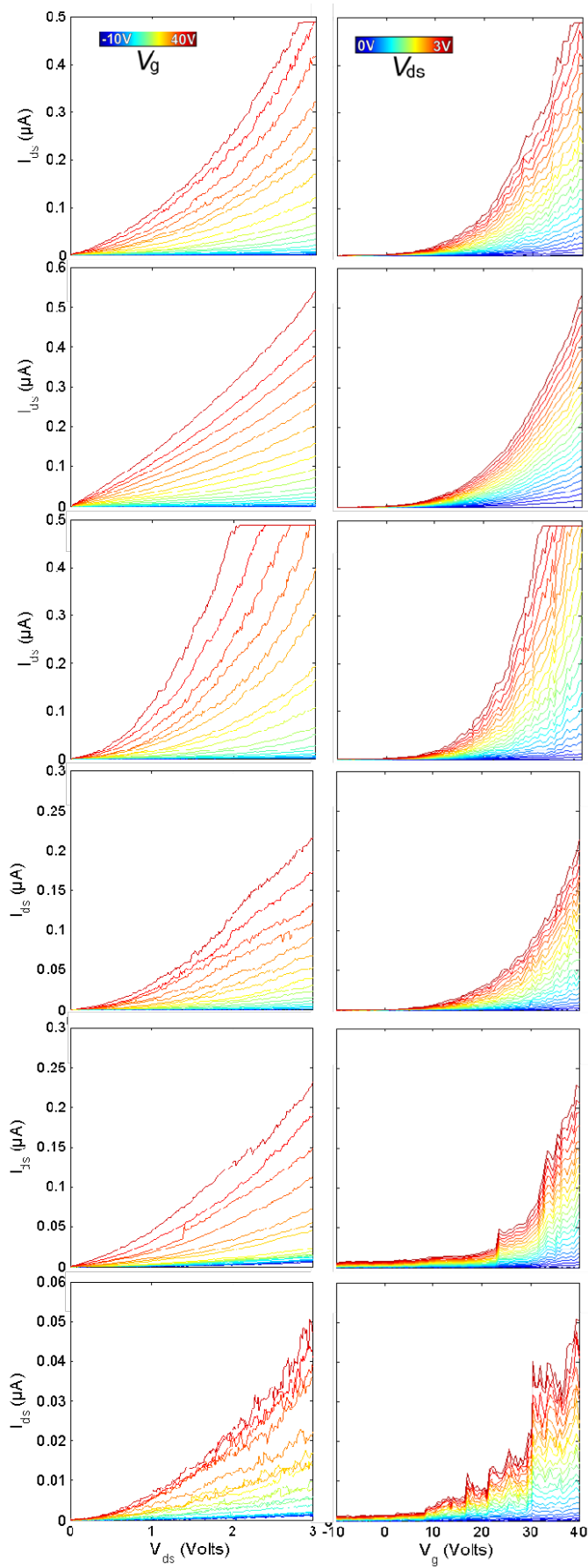

**Figure S7.** $I_{ds}$-$V_{ds}$ and $I_{ds}$-$V_g$ characteristics measured for six different FET devices based laser-fabricated MoS$_2$ monolayers. The $I_{ds}$-$V_{ds}$ characteristics have been measured for different gate voltages ranging from -10 V (blue traces) to +40 V (red traces). The $I_{ds}$-$V_g$ characteristics have been measured for different drain source voltages ranging from -0 V (blue traces) to +3 V (red traces).

**References of the supplementary material**